\documentclass[journal]{IEEEtran}
\usepackage{float}
\usepackage{cite}
\usepackage{amsmath, amssymb}
\usepackage{tabularx}
\usepackage{ dsfont }
\usepackage[short]{optidef}
\usepackage[hyphens]{url}
\usepackage{hyperref}
\usepackage{cancel}
\usepackage{ mathrsfs }
\usepackage{graphicx}
\usepackage{soul,xcolor}
\usepackage{todonotes}
\usepackage{stackengine}
\usepackage{scalerel}
\usepackage{enumitem}
\usepackage{nomencl}
\usepackage{booktabs}

\usepackage[subtle]{savetrees}
\allowdisplaybreaks[4]
\renewcommand{\baselinestretch}{0.93}
\usepackage{amsthm}
\theoremstyle{remark}
\newtheorem{remark}{\textit{Remark}}

\definecolor{Light}{gray}{0.85}

\begin{document}
\title{Real-time Grid and DER Co-simulation Platform for Validating Large-scale DER Control Schemes}

\author{ Adil Khurram,~\IEEEmembership{Student Member,~IEEE,} Mahraz Amini,~\IEEEmembership{Member,~IEEE,}, Luis A. Duffaut Espinosa,\\ Paul D. H. Hines, Mads Almassalkhi,~\IEEEmembership{Senior Members,~IEEE}

\thanks{This work was supported by the U.S. Department of Energy's Advanced Research Projects Agency - Energy (ARPA-E) award DE-AR0000694 and NSF grant CMMI-1839387.
P.~Hines and M.~Almassalkhi are co-founders of startup Packetized Energy, which seeks to bring to market a commercially viable version of Packetized Energy Management.}
\thanks{A.~Khurram, L.~A.~D.~Espinosa, P.~D.~H.~Hines and M.~Almassalkhi are with the Department of Electrical and Biomedical Engineering, University of Vermont, Burlington,VT, 05405 USA e-mail: \{akhurram,lduffaut,phines,malmassa\}@uvm.edu. M.~Amini is with National Grid, MA, USA e-mail: mahraz.amini@nationalgrid.com}
}

\markboth{IEEE Transactions on Smart Grid}{}%
\maketitle

\begin{abstract}
Distributed energy resources (DERs) such as responsive loads and energy storage systems are valuable resources available to grid operators for balancing supply-demand mismatches via load coordination. However, consumer acceptance of load coordination schemes depends on ensuring quality of service (QoS), which embodies device-level constraints. Since each device has its own internal energy state, the effect of QoS on the fleet can be cast as fleet-wide energy limits within which the aggregate ``state of charge'' (SoC) must be actively maintained. This requires coordination of DERs that is cognizant of the SoC, responsive to grid conditions, and depends on fast communication networks. To that effect, this paper presents a novel real-time grid-and-DER co-simulation platform for validating advanced DER coordination schemes and characterizing the capability of such a DER fleet. In particular, we present how the co-simulation platform is suitable for: $i$) testing real-time performance of a large fleet of DERs in delivering advanced grid services, including frequency regulation; $ii$) online state estimation to characterize the corresponding SoC of a large fleet of DERs;  and $iii$) incorporating practical limitations of DERs and communications and analyzing the effects on fleet-wide performance. To illustrate these benefits of the presented grid-DER co-simulation platform, we employ the advanced DER coordination scheme called packetized energy management (PEM), which is a novel device-driven, asynchronous, and randomizing control paradigm for DERs. A fleet of thousands of PEM-enabled DERs are then added to a realistic and dynamical model of the Vermont transmission system to complete validation of the co-simulation platform.
\end{abstract}

\IEEEpeerreviewmaketitle

\section{Introduction}
The drive to reduce greenhouse gas emissions and declining capital costs of distributed renewable generation is causing rapid increase in wind and solar generation capacity. However, despite their low emissions profile, wind and solar power generation vary rapidly in time, motivating the need for additional balancing resources~\cite{AminiPSCC2018,schweppe1980homeostatic,Morgan:1979,malhame1990AAP}. The availability of internet connected and controllable distributed energy resources (DERs) have made it possible to balance mismatches between generation and load via DER coordination. DERs such as electric water heaters (EWHs) and energy storage systems (ESS) can be coordinated to provide grid services without any degradation in quality of service (QoS) to the customer \cite{Callaway2009ECM, Meyn2015TAC, Mathieu:2013tt, Luminita2014IFAC, almassalkhi2018BookChapter,DuffautEspinosa:2018PSCC, Duffaut-et-al_2019a,Duffaut-et-al_2019b,  malhame1990AAP}. The capability to aggregate and control DERs to provide services to the grid with QoS guarantees is called demand dispatch.

The main ideas underlying modern demand dispatch have existed for decades\cite{schweppe1980homeostatic,Morgan:1979, malhame1990AAP}. The pioneering work focusing on direct load control of electric water heaters \cite{malhame1990AAP} provided analytical tools to evaluate the effect of control strategies such as cold load pick-up on a group of EWHs. Since then, several DER coordination schemes have been proposed and validated in simulations. These schemes either assume full control over DER's operating state and knowledge of its state of charge (SoC) in a centralized coordinator~\cite{Callaway2009ECM} or influence power consumption of DERs indirectly by transmitting a control command at regular intervals in a distributed coordinator~\cite{Mathieu:2013tt,Luminita2014IFAC, Meyn2015TAC,almassalkhi2018BookChapter,DuffautEspinosa:2018PSCC, Duffaut-et-al_2019a,Duffaut-et-al_2019b}. Packetized energy management (PEM) is one such distributed load coordination scheme in which DERs request the PEM coordinator to consume power which can be accepted or rejected as required~\cite{almassalkhi2018BookChapter,DuffautEspinosa:2018PSCC, Duffaut-et-al_2019a,Duffaut-et-al_2019b}. Distributed coordinators, like PEM, assume access to limited DER information but employ model based state estimation schemes to estimate the SoC of the fleet. Nonetheless, to achieve demand dispatch communication is required between DERs and coordinators that gives rise to cyber-physical systems (CPS).

Recent advances in sensing and connectivity have led to widespread adoption of cyber-physical systems as a framework in which to incorporate, model, and analyze the interactions between the cyber (e.g., communication networks) and the physical (e.g., grid)~\cite{Serpanos2018Computer, Xin2015TSG, Ayar2017IET, Korukonda2018IET}. In this regard, CPS test-beds have been developed within the smart grid area that consists of a combination of physical and simulated components~\cite{NREL_CPS:EUROCON2019,Hahn:TSG2013, UIUC_CPS:Informs2009} to study stability and performance of grid control algorithms in the presence of faults, measurement noise, latency in communication and malicious data injection. Electric power generation and distribution systems are either simulated using computer based softwares such as PowerWorld~\cite{UIUC_CPS:Informs2009} or by means of specially designed hardware-based grid simulators e.g. OPAL-RT~\cite{NREL_CPS:EUROCON2019} and real-time digital simulator~\cite{Hahn:TSG2013}. Physical components like relays, circuit breakers, inverters, PV emulators etc. that operate at high-voltages are integrated with the simulated electric grid by means of appropriate power converters~\cite{NREL_CPS:EUROCON2019}. Special attention is given to implement standard communication protocols used for data transfer and sending control commands to recreate real-world scenarios.

However, these detailed test-beds are particularly designed for the purpose of transient stability analysis and evaluating the cyber-security performance, for example, studying the effect of a malicious breaker-trip that can cause harmful oscillations and potentially unbalance the power system~\cite{Hahn:TSG2013}. For this purpose, only a limited number of physical devices are sufficient. Load coordination schemes, on the other hand, require thousands of DERs to create a net-effect reasonable for providing grid services.

The literature on advanced DER coordination has mostly been in the form of theoretical schemes and simulations that ignore either the grid or the real-time effects of DERs and communication networks~\cite{Meyn2015TAC, Mathieu:2013tt}. These load-coordination schemes are designed to operate using the customer's internet connection and their aggregate behavior has been extensively studied and subsequently modeled~\cite{Mathieu:2013tt,Luminita2014IFAC, Meyn2015TAC,almassalkhi2018BookChapter, DuffautEspinosa:2018PSCC, Duffaut-et-al_2019a, Duffaut-et-al_2019b}. Although physical limitations such as communication delays have been incorporated in Matlab based simulated environments~\cite{Ledva:2017ws}. However, there is a pressing need to have access to realistic and real-time CPS test-beds in order to convincingly evaluate the potential of large fleets of DERs to provide grid services in a laboratory setting since thousands of DERs are not readily available in the field for validation. Furthermore, the CPS test-bed should be able to capture the limitations associated with real-world deployments.

Towards this goal, this paper presents a real-time grid-and-DER co-simulation platform for the large scale validation of DER coordination schemes on realistic and dynamic transmission systems. The proposed platform is particularly useful for $i$) online state estimation to characterize the corresponding SoC of a large fleet of DERs; $ii$) testing real-time performance of a large fleet of DERs in delivering advanced grid services, including frequency regulation; and $iii$) incorporating practical limitations of DERs and communications as well as analyzing the effects on fleet-wide performance. Furthermore, the co-simulator is agnostic to the type of DER coordination scheme. Illustrative results are presented for packetized energy management, a bi-directional DER-to-Coordinator demand dispatch scheme developed in~\cite{almassalkhi2018BookChapter,DuffautEspinosa:2018PSCC, Duffaut-et-al_2019a, Duffaut-et-al_2019b} for which preliminary validation is carried out in~\cite{Amini2019PESGM} along with the early field-trial testing with customer-owned electric water heaters in~\cite{Adil2019ISGT}. This paper extends the author's earlier work by incorporating real-time state estimation and practical limitations as described above.

The remainder of the paper is organized as follows. In section~\ref{sec:RTsim}, the grid-and-DER co-simulation platform is presented. Section~\ref{sec:PEM_prelim} describes packetized energy management of DERs.  Sections~\ref{sec:agg_valid},~\ref{sec:grid_service} and~\ref{sec:delay} subsequently deal with the implementation of advanced aggregate models for control and estimation, validation of grid services and identification of practical limitations associated with demand dispatch. Finally, section~\ref{sec:concl} concludes the paper.

\section{Real-time grid-and-DER co-simulation platform} \label{sec:RTsim}
This section describes the real-time cyber enabled platform developed in this work. Real-time herein refers to the orders of tens of milliseconds. The proposed validation platform consists of four components, ($i$) grid simulator, ($ii$) DER emulator, ($iii$) communication channel, ($iv$) DER coordinator or aggregator. The over-arching goal of this platform is to provide means to study and demonstrate the effects of large scale demand dispatch on the grid in a realistic setting as well as identify potential DER related challenges associated with such deployment.

\begin{figure}[h]
    \centering
    \includegraphics[width=1\columnwidth]{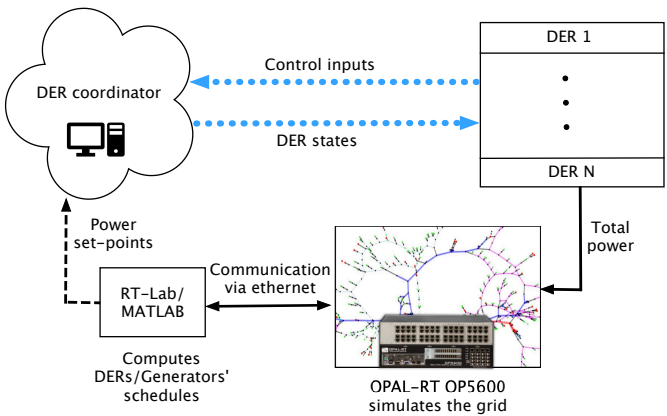}
    \caption{Cyber-physical platform overview: The transmission grid is simulated on OP5600 and AGC based dispatch is realized on a host PC that dispatches demand. The packetized load is emulated on a high performance PC that is aggregated by a python based server. 
    }
    \label{fig:fullsys_overview}
\end{figure}
\subsection{Grid Simulator}
The electric grid is simulated using the OP5600 real-time digital simulator from OPAL-RT. The OP5600 has a multi-core processor along with digital and analog I/O with the capability of interfacing to a network of PCs in order to simulate large models in real-time with a small time-step of integration which is on the order of milliseconds. The RT-Lab software allows the communication between a host PC and the target (OP5600) simulator such that a real-time physical model can be simulated on the OP5600 while the controller would be executed on the PC where an operator could make adjustments when necessary.

``ePHASORSIM" is a tool developed by OPAL-RT to offer dynamic simulation of power systems in order to conduct power system studies and test control schemes. The grid is modeled with a standard equivalent positive-sequence single-phase constant-power AC model in ePHASORSIM. In this paper, the electric grid is based on the Vermont Electric Power Company's (VELCO) transmission system and its details are further discussed in section \ref{Subsec:VELCOtestcase}. Furthermore, RT-LAB and ePHASORSIM can be interfaced with Simulink, which is used to develop controls for the power system. The OPAL-RT blockset for Simulink allows a section of the Simulink block diagram to be executed in real-time on the OP5600 and the controls can be executed asynchronously on the PC with the ability to accept user inputs when necessary.
\subsection{DER coordinator}
\begin{figure}[ht]
    \centering
    \includegraphics[width=\columnwidth]{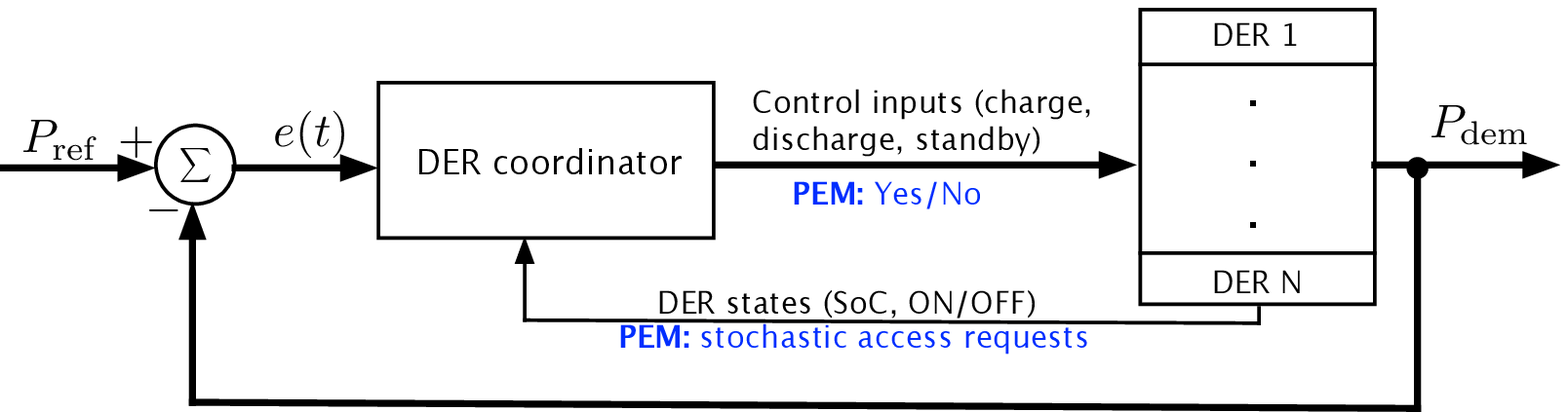}
    \caption{Closed-loop feedback system for demand dispatch to track $P_{\rm{ref}}$ based on the aggregate net-load measured by the coordinator and control inputs to DERs.}
    \label{fig:PEMschematic}
\end{figure}

The coordinator managing DERs sends control signals to the fleet depending upon the power reference $P_{\text{ref}}$ provided by the utility and or aggregator, the total power consumption of the fleet $P_{\text{dem}}$ and measurements regarding DER's states. The block diagram in Fig.~\ref{fig:PEMschematic} represents a general DER coordination scheme where the control signal can be ON/OFF commands for individual DERs or an ON/OFF switching fraction that is broadcast to the entire fleet~\cite{Luminita2014IFAC}. In case of PEM~\cite{almassalkhi2018BookChapter}, it is simply a YES/NO response to a DER's request which is described in detail in the next section.

The DER coordinator is implemented in Python using an open-source event-driven networking engine called \texttt{Twisted}. Specifically, a TCP server listens for Hypertext Transfer Protocol (HTTP) messages from the emulated DERs. Updates regarding DER's states and control inputs such as ON/OFF commands, are received or transmitted via the HTTP-POST and GET methods respectively using Python's \texttt{requests} module. Different demand dispatch schemes can, therefore, be implemented in this setup, such as but not limited to, the load control of \cite{Luminita2014IFAC} or the PEM based demand dispatch~\cite{almassalkhi2018BookChapter}. 

\subsection{Emulation of DERs}
DER logic is written in C/C++ based on a simplified first order state of charge dynamics.
A DER is then initialized as a thread that executes its logic as a background process in the processor of a high performance computer. Using this setup, thousands of DERs can be emulated by simply spawning a thread for each DER in the processor's memory. Each thread then emulates a DER equipped with its own clock. The key difference between this implementation and other Matlab based simulations is that each DER executes its logic independently of other DERs. Finally, to enable demand dispatch, communication between the DERs and the DER coordinator is achieved over the internet via the HTTP requests mechanism. Demand dispatch via internet enabled DERs is becoming more common, such as those presented in~\cite{Adil2019ISGT}, which is accurately represented in this platform. Although, the results presented in this paper are for PEM, however, the DER implementation is flexible enough to allow any type of demand dispatch.

\subsection{Communication channel}
As mentioned earlier, the communication between DERs and the coordinator occurs by means of HTTP requests over the internet. This type of implementation allows one to identify practical limitations associated with real-world deployment that includes but not limited to communication latency, packet loss and loss of internet connection at the. The effect of communication latency is specifically studied in section~\ref{sec:delay}. The next remark highlights this fact.

\begin{remark}
    Both the DER emulator and the coordinator are executed on the same PC, however, it is important to highlight that the coordinator and each of the emulated DER are executed as a separated thread in the computer's memory. In short this means that decisions made by the coordinator will not affect the requesting DERs instantaneously rather control inputs to DERs must pass through the communication channel that incorporates delays as well as dynamics of the channel itself. This is essential to ensure asynchronism is maintained which has a significant impact in real-world DERs but usually ignored in Matlab based simulation studies.
\end{remark}

\section{Packetized energy management preliminaries} \label{sec:PEM_prelim}
Packetized energy management (PEM) is a device driven demand dispatch scheme that is used to coordinate DERs in a way that allows groups of DERs to provide a wide range of different grid services, including peak reduction, energy price arbitrage and even frequency regulation. This section discusses preliminaries of PEM.

\subsection{Device-driven coordination of DERs} \label{subsec:PEM_micro}
A DER with SoC $x$, operates within a deadband $[\underline{x}, \overline{x}]$ around a user-specified setpoint $x_{\text{set}}$ where $\underline{x}$ and $\overline{x}$ are the lower and upper bounds of the deadband. Then, based on their SoC, PEM enabled DERs can be in one of the four logical modes i.e. ($i$) charge, ($ii$) discharge, ($iii$) standby and ($iv$) opt-out \cite{almassalkhi2018BookChapter}. In the standby mode with $x \in [\underline{x}, \overline{x}]$, DERs requests the aggregator to either consume power in the charge mode from the grid or inject power into the grid in discharge mode which is called a packet request. If a charging (discharging) packet request is accepted, then the DER transitions to charge (discharge) mode for a specified period of time called charging (discharging) packet length and is denoted as $\delta_{1}$ ($\delta_{-1}$). Furthermore, the opt-out mode is included in PEM that ensures QoS by allowing DERs to temporarily opt-out of PEM if $x < \underline{x}$ and charge until SoC has sufficiently recovered. DERs can also opt-out due to excessive energy ($x > \overline{x}$). The request mechanism is described next.

The probability that a DER in ``standby'' mode  with SoC $x[k]$ at time $k$ requests a charging packet is designed to be the following cumulative exponential distribution function,
\begin{align}
g_{\mu}(x[k]) &= 1 - e^{-\mu(x[k])\Delta t}\label{eq:PEM_stoch_req},\\
\mu(x[k]) &= \left\{\begin{matrix*}[l]
0, & \text{if } x[k]\ge \overline{x}\\
m_R(\frac{\overline{x} - x[k]}{x[k] - \underline{x}}) (\frac{x_{\text{set}} - \underline{x}[k]}{\overline{x}[k] - x_{\text{set}}}), & \text{if } x[k]\in (\underline{x}, \overline{x})\\
\infty, & \text{if } x[k]\le \underline{x} \label{eq:PEM_mu_chg}
\end{matrix*}\right.
\end{align}
where, $\mu(x[k])$ is called the rate parameter, $m_{R}$ is a design parameter called the mean time to request. The discharge probability of request can be obtained in a similar manner~\cite{almassalkhi2018BookChapter}. The probability of request function ($g_\mu$) is designed to ensure that a DER with lower SoC has a greater probability of making a request as compared to the one with higher energy.

The SoC of the individual DER, evolves according to the first order difference equation,
\begin{align}
    x[k+1] = x[k] + \Delta t \left( \phi_{\zeta}P_{\zeta}\zeta[k]  + \phi_{\text{s}}[k]x[k] + \phi_{\text{u}}[k] \right),\label{eq:TCLmodelDT}
\end{align}
where $x$ is the SoC, $\zeta \in \{-1,0,1\}$ is the state of the thermodynamic switch, $\phi_\zeta$ is the charging or discharging efficiency, $\phi_{\text{s}}$ represents the standing losses and $\phi_{\text{u}}$ models the effect of end-use consumption. When $\zeta = 1$ ($\zeta = -1$) then the DER is charging (discharging) at the power transfer rate $P_{1}$ ($P_{-1}$)kW, efficiency $\eta_{1}$ ($\eta_{-1}$) and the DER is in standby mode when $\zeta=0$.

Consider thermostatically controlled loads such as electric water heaters that can only charge, that is $\zeta \in \{ 0, 1\}$, let $\Theta_{\text{EWH}} = \{P_{\zeta},L,x_{\text{set}}, [\underline{x}, \overline{x}]\}$ be the set of parameters. Then in~(\ref{eq:TCLmodelDT}$), \phi_{\zeta} = (\eta_{\zeta} P_{\zeta})(c \rho L)^{-1}$ where $\eta_{\zeta}$ is the efficiency, $c = 4.186$ $(\text{kJ})(\text{kg} ^\circ C)^{-1}$ is the specific heat constant, $\rho = 0.990$  $\text{kg}\text{L}^{-1}$ is the density of water when close to $50$ $^\circ C$ and $L$ is the tank size in liters. Furthermore, $\phi_{\text{s}}[k] = (x[k]-x_{\text{a}})(x[k]\tau)^{-1}$ and $\phi_{\text{u}}[k]=Q[k] (c \rho L)^{-1}$ where, $x_{\text{a}}=21^\circ C$ is the ambient temperature, $\tau = 150\times 3600$ seconds is the standing loss time constant to ambient temperature and $Q[k]$ is the heat loss from the tank due to customer driven water usage and is modeled as Poisson random pulses~\cite{Duffaut-et-al_2019a}. Similarly, energy storage systems (ESSs) with capacity $\mathcal{C}$ in kWh can both charge and discharge with efficiencies $\phi_{1} = \eta_{1}$ and $\phi_{-1} = \eta_{-1}$ respectively. End-use consumption and standing losses for ESSs are assumed to be negligible, $\phi_{\text{s}} = \phi_{\text{u}}=0$ resulting in the parameter set $\Theta_{\text{ESS}} = \{P_{\zeta},\mathcal{C}, \eta_{\zeta}, x_{\text{set}}, [\underline{x}, \overline{x}]\}$. Next, state-bin transition model is briefly described that is used to capture the aggregate dynamics of the fleet~\cite{Duffaut-et-al_2019a}.

\subsection{Macromodel for aggregate population}
For DERs, the Law of Large Numbers admits a state bin transition model which is referred to as the macromodel in this paper. Macromodel is a Markovian model that accurately approximates the average behaviour of a sufficiently large fleet. Consider first a DER in charge mode ($\zeta=1$) whose dynamic state $x[k]$ evolves in the interval $[\underline{x},\overline{x}]$. This continuous interval $[\underline{x}, \overline{x}]$ is divided into $n_{\text{b}}$ discrete states or bins and collected in the set $\mathcal{X}_{\text{c}}:=\{x^{\text{c}}_1, \dots, x^{\text{c}}_{n_{\text{b}}} \}$. Henceforth, the sub-scripts (super-scripts) c, d, and sb correspond to charge, discharge and standby modes respectively. Let $X_{k}, k\ge 0$ be a random variable that takes on values in $\mathcal{X}_{\text{c}}$ and describes the dynamic state of the charging DER at time $k$. $X_k$ transitions from state $x^{\text{c}}_i \in \mathcal{X}_{\text{c}}$ to $x^{\text{c}}_{i+1}$ with probability $m^{\text{c}}_{i(i+1)}$ and is called the transition probability. The transition probability for each of the states are obtained from (\ref{eq:PEM_stoch_req}) as outlined in \cite{Duffaut-et-al_2019a}. Further define $q_{\text{c}}[k]:= (q_1^{\text{c}}, \dots, q_{n_{\text{b}}}^{\text{c}})^\top$ as the probability mass function (PMF) of $X_k$ where each $q_i^{\text{c}}$ is the probability that $X_k = x^{\text{c}}_i$. For a fleet of DERs, $q_i^{\text{c}}$ is the total proportion of DERs in charge mode having the dynamic state $x_i^{\text{c}}$. The complete state space and PMF can now be constructed as $\mathcal{X} = \mathcal{X}_{\text{c}} \cup \mathcal{X}_{\text{sb}} \cup \mathcal{X}_{\text{d}}$ and $q=(q_{\text{c}}^\top, q_{\text{sb}}^\top, q_{\text{d}}^\top)$ respectively. Both $\mathcal{X}_{\text{sb}}$ and $\mathcal{X}_{\text{d}}$ are identical copies of $\mathcal{X}_{\text{c}}$. The transition probabilities $m^{\text{sb}}_{(i-1)i}$ correspond to standby mode and define backward transition from states $x^{\text{sb}}_i$ to $x^{\text{sb}}_{i-1}$. Similarly, $m^{\text{d}}_{(i-1)i}$ are transition probabilities corresponding to discharge mode. 

Thus, over any given interval $k$, the DER coordinator receives a number of charging and discharging requests. The coordinator then determines the proportion of charging and discharging requests to be accepted denoted as $\beta_{\text{c}}[k]$ and $\beta_{\text{d}}[k]$ respectively. This effect is achieved in the macromodel by transitioning the DERs from standby to corresponding charging and discharging states. Moreover, a timer tracks the total number of expiring packets $\beta^-_{\text{c}}[k]$ and $\beta^-_{\text{d}}[k]$ for past charging and discharging requests respectively to ensure transition to standby states. The resulting dynamic is nonlinear and can be written as
\begin{align}
    q[k+1] = f(\beta[k], \beta^-[k], q[k]),\quad y[k] = h(q[k]) \label{eq:PEM_nonlin_dyn}
\end{align}
where, $\beta[k] = (\beta_{\text{c}}[k], \beta_{\text{d}}[k])$, $\beta^-[k] = (\beta^-_{\text{c}}[k], \beta^-_{\text{d}}[k])$, $f$ is a non-linear mapping, $f:\mathbb{R}^{4+3n_{\text{b}}} \rightarrow \mathbb{R}^{3n_{\text{b}}}$. Finally, $y\in \mathbb{R}^3$ is the output vector and $h(q[k]) = (h_{\text{dem}}(.), h_{\text{req,c}}(.), h_{\text{req,d}}(.))^\top$, where $h_{\text{dem}}, h_{\text{req,c}}, h_{\text{req,d}}$ are one dimensional mapping from state to total power demand, total number of charging requests and total number of discharging requests respectively. For full details of the macromodel, the reader is referred to \cite{Duffaut-et-al_2019a}. 

\subsection{Baseline power consumption of a DER fleet}
The baseline power consumption of a fleet of DERs of the same load type is the minimum power signal ($P_{\text{nom}}$) for which the quality of service is guaranteed by PEM \cite{DuffautEspinosa:2018PSCC}. QoS is defined in terms of the average SoC of the fleet. For the case of electric water heaters, SoC is the average tank temperature of the fleet. If the average temperature is close to the setpoint, then we say that QoS is satisfied. Macromodel is used to obtain the baseline or nominal power consumption of the fleet and is obtained by solving the following optimization problem \cite{Duffaut-et-al_2019b},
\begin{subequations}   \label{eq:BetaNom}
	\begin{align}
	& \hspace*{-0.6in} P^*_{\text{nom}} := \min_{\beta_{\text{c}},\beta_{\text{d}} \in [0,1]} \;\; \sum_{i=1}^n h_{\text{dem}}(q^*)  \;\;\;\; \mbox{subject to} \label{eq:MinPowerBetaNom}\\
		\qquad & q^*   = f{(\beta[k],\beta^-[k], q^*)}, \label{eq:InvDistrBetaNom}\\
	\qquad &(q^*)^\top \, \chi \ge x_{\text{set}}, \label{eq:SetPointEqBetaNom}
	\end{align}
\end{subequations}
where, $x_{\text{set}}$ is the setpoint and $\chi\in \mathbb{R}^{3n_{\text{b}}}$ consists of SoC associated with the state space  $\mathcal{X}$. The solution of the above optimization problem provides the total proportion of charging and/or discharging requests to be accepted ($\beta_c[k], \beta_d[k]$) that directly corresponds to a steady state distribution $q^*$ in the macromodel and hence the baseline power consumption is obtained from $P_{\text{nom}} = h_{\text{dem}}(q^*)$. The next section presents details regarding implementation of PEM enabled DERs on the grid-and-DER co-simulation platform and uses the macromodel to estimate SoC of the fleet.

\subsection{Implementation of PEM on simulator} \label{subsec:PEM_implement}
The emulated DERs under PEM are executed as a thread in the processor's memory. Each DER then samples from the probability of request curve to determine if a request to consume a packet (charging or discharging but not both) is to be made to the coordinator. In case a request is made, the DER sends that request and then waits for the response. Requests are sent over the internet using HTTP and Python's \texttt{requests} module. 

The coordinator, after receiving the request, either accepts or rejects it depending upon the total power consumption of the fleet and the power reference so that the tracking error is zero. That is, for the tracking error $P_{\text{error}}[k]:= P_{\text{ref}}[k] - P_{\text{dem}}[k]$ a charge packet request with rated power $P_{1}$ is accepted if $P_{\text{error}}[k] + P_{1} \le 0$ and rejected otherwise. On the other hand, a discharge packet with rated power $P_{-1}$ is accepted only if $P_{\text{error}}[k] < 0 $ and $P_{\text{error}}[k] + P_{-1} \ge 0$. It should be emphasized here that the coordinator is blind to the requesting DER's identity while making a decision to ensure privacy.

Each DER is equipped with a local clock to keep track of time elapsed since it started consuming a packet. As a result, asynchronism is ensured that is absent in standard software based simulations. In addition to the DER's local timer, the coordinator also associates a timer with each accepted packet request. This is achieved using Python's \texttt{threading} module. Timers allow the coordinator to keep track of the expiring packets and enables tighter tracking performance as shown in section~\ref{sec:delay}. It should also be noted here that the coordinator's ability to influence its decisions without explicit measurements such as DER's power is made possible only due to the PEM's unique request-response mechanism that acts as a natural feedback. Furthermore, this feature is an advancement over the author's earlier implementation of the coordinator in~\cite{Adil2019ISGT}.

\subsubsection{Illustration with real-time emulated DERs}
\begin{figure}[ht]
    \centering
    \includegraphics[width=\columnwidth]{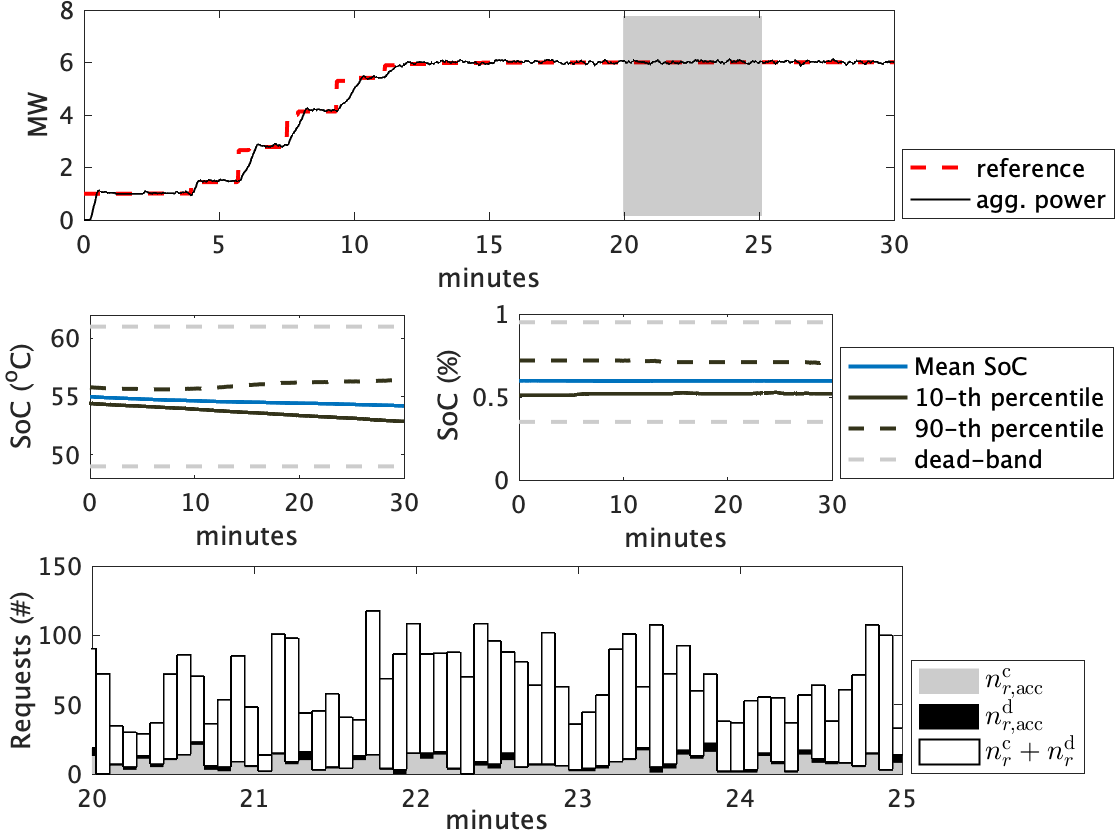}
    \caption{This figure shows the tracking performance as well as the SoC of a fleet of heterogeneous DERs. The fleet consists of $4,900$ EWHs and $1,150$ ESSs emulated in real-time using the platform developed in this work. Two plots in the middle show the mean, $10$-percentile and $90$-th percentiles of the SoC of EWHs and ESS. Finally, the bottom plot shows the total number of accepted charging requests ($n_{r, \text{acc}}^{\text{c}}$) in black, total number of accepted discharging requests ($n_{r, \text{acc}}^{\text{d}}$) in grey color and aggregate charging and discharging requests made $n_{r}^{\text{c}} + n_{r}^{\text{d}}$.}
    \label{fig:EWH_ESS_Het}
\end{figure}

Consider a heterogeneous fleet of diverse DERs consisting of $4,900$ EWHs and $1,150$ ESSs emulated in real-time using the developed simulator. The EWH parameters are $\Theta_{\text{EWH}} = \{\mathcal{P}_{\zeta}^{\text{EWH}},\mathcal{L},52, [48.9, 55.1]\}$, $\mathcal{P}_{\zeta}^{\text{EWH}} \sim \mathcal{N}(4.5, 0.25)$,  $\mathcal{L} \sim \mathcal{N}(200, 40)$ where $\mathcal{N}(\mu, \sigma)$ represents normal distribution with mean $\mu$ and standard deviation $\sigma$. Similarly, $\Theta_{\text{ESS}} = \{\mathcal{P}_{\zeta}^{\text{ESS}},\mathcal{H},100, 75, [55, 95]\}$, $\mathcal{P}_{\zeta}^{\text{ESS}} \sim \mathcal{N}(5, 0.5)$,  $\mathcal{H} \sim \mathcal{N}(13.5, 1)$. Grid is neglected in this example and $P_{\zeta}$, $\eta_{\zeta}$ for for charge and discharge modes are the same in ESSs. The fleet is coordinated by the aforementioned server that accepts or rejects requests to track a reference signal that varies from $1$ MW to $6$ MW in steps of $1$ MW as shown in Fig.~\ref{fig:EWH_ESS_Het}.

The average SoC of EWHs and ESS is plotted in Fig.~\ref{fig:EWH_ESS_Het} along with the $10$-th and $90$-th percentiles of the population during tracking. This shows that the SoC of the fleet remains close to the setpoint during the experiment. Furthermore, the total number of accepted charging ($n_{r, \text{acc}}^{\text{c}}$) and discharging ($n_{r, \text{acc}}^{\text{d}}$) requests are also shown as the shaded region in a bar plot at the bottom of Fig.~\ref{fig:EWH_ESS_Het}. Also included in the bar plot is the total number of charging and discharging requests ($n_{r}^{\text{c}} + n_{r}^{\text{d}}$) made by the EWHs and ESSs respectively. 
\section{Implementation of advanced aggregated models} \label{sec:agg_valid}
The objective of this section is to demonstrate that the cyber enabled simulator provides a realistic environment for testing advanced algorithms, such as the PEM macromodel described in the previous section, before real-world deployment. 
\subsection{State of charge}
DERs are limited by the physical device constraints and can only be charged or discharged for a limited time. SoC of a fleet of DERs quantifies the flexibility available to the coordinator and is useful for making informed decisions. Let $x_i[k]$ be the SoC of the $i$-th DER, then the SoC of the fleet denoted as $z[k]$ is defined as the average value of $x_i[k]$ normalized with respect to higher ($\overline{x}$) and lower ($\underline{x}$) limits  as follows,
\begin{align}
    z[k] := \frac{ \sum_{i=1}^N \frac{ x_i[k]}{N} - \underline{x}}{\overline{x} - \underline{x}}. \label{eq:soc_def}
\end{align}
\begin{remark}
Clearly, SoC ($z[k]$) in (\ref{eq:soc_def}) requires individual state $x_i[k]$ to be measured. However, PEM does not measure each state, therefore, an estimate of $z[k]$ is required. For this purpose, the extended Kalman filter (EKF) based on the macromodel has been proposed in \cite{DuffautEspinosa:2018PSCC} and can be used to estimate the SoC of the fleet. In this paper, it is assumed that groups of DERs of the same load type are aggregated and a SoC can be associated with each of those groups.
\end{remark}
\subsection{State of charge limits}
In PEM, SoC of the fleet achieves it maximum value after authorizing all charging requests and rejecting all discharging requests for a long time. In terms of the macromodel, this is achieved by setting $\beta_{\text{c}}[k] = 1$ and $\beta_{\text{d}}[k]=0$. By fixing the control signal $\beta[k]=(1,0)$ and the timer dynamics, \eqref{eq:PEM_nonlin_dyn} is reduced to an aperiodic, irreducible Markov chain which is guaranteed to have a stationary distribution. The maximum SoC limit, $\overline{z}$, is then the average SoC of the fleet corresponding to that distribution. Similarly, the minimum SoC limit is obtained by rejecting all charging requests ($\beta_{\text{c}}[k]=0$) and accepting all discharging requests ($\beta_{\text{d}}[k]=1$). The resulting stationary distribution of \eqref{eq:PEM_nonlin_dyn} gives the minimum limit $\underline{z}$. The state estimation procedure using EKF is presented next.

\subsection{State estimation}
The Extended Kalman filter is used for state estimation (SE) that uses as measurements, $i)$ the total requests, $ii)$ aggregated power demand of the fleet and $iii)$ opt-out rates, which are available to the coordinator. The observability of PEM macromodel has previously been discussed in \cite{DuffautEspinosa:2018PSCC} where an EKF is designed. The EKF formulation is provided here for completeness.
\begin{enumerate}
    \item Measurement update:
    \begin{align*}
    K[k] &= \text{Cov}[k|k-1] C^\top (C \text{Cov}[k|k-1]C^\top + \mathcal{Q}_1[k])^{-1}\\
    \text{Cov}[k|k] &= \text{Cov}[k|k-1] - K[k]C \text{Cov}[k|k-1]\\
    \hat{q}[k|k] &= \hat{q}[k|k-1] + K[k](y[k] - C \hat{q}[k|k-1])
    \end{align*}
    \item Time update
    \begin{align*}
    \text{Cov}[k+1|k]&= A_{u[k]}\text{Cov}[k|k]A_{u[k]}^\top + \mathcal{Q}_2[k]\\
    \hat{q}[k+1|k] &= A_{u[k]}\hat{q}[k|k]
    \end{align*}
\end{enumerate}
where $\hat{q}[k|k-1]$ is the predicted state of the system, $\hat{q}[k|k]$ is the updated estimate of the system, $A_{u[k]}$ is the system matrix obtained by linearizing (\ref{eq:PEM_nonlin_dyn}) for the input $u[k] = (\beta[k], \beta^-[k])^\top$, $C$ is the output matrix obtained by linearizing the output equation $y[k]$ from (\ref{eq:PEM_nonlin_dyn}). Furthermore, $\text{Cov}[k|k-1]$ and $\text{Cov}[k|k]$ denotes the predicted system covariance and updated system covariance respectively, $\mathcal{Q}_1$ and $\mathcal{Q}_2$ are measurement and process noise covariance respectively.

\subsection{Illustrating SE on a fleet of EWHs}
\begin{figure}[ht]
    \centering
    \includegraphics[width=\columnwidth]{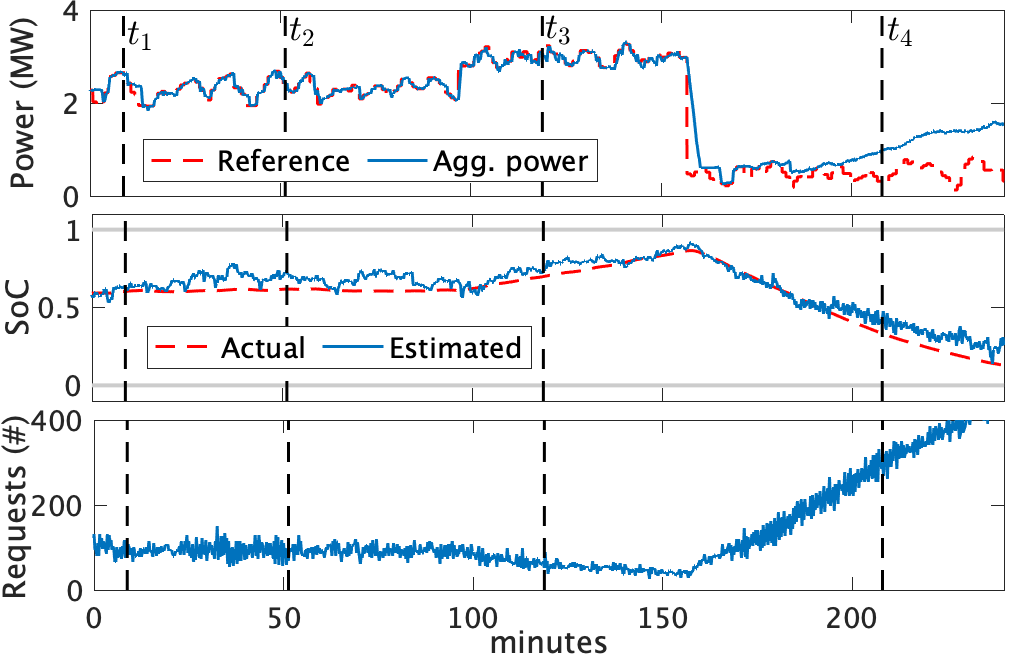}
    \caption{Online estimation of the SoC (center) of DER aggregation consisting of $2,000$ EWHs from measurements of total power consumption (top), number of request (bottom) and opt-outs only. DERs  are shown to charge and discharge while tracking a scaled and shifted AGC signal.}
    \label{fig:PVB_EKF_validation}
\end{figure}

\begin{figure}[ht]
    \centering
    \includegraphics[width=\columnwidth]{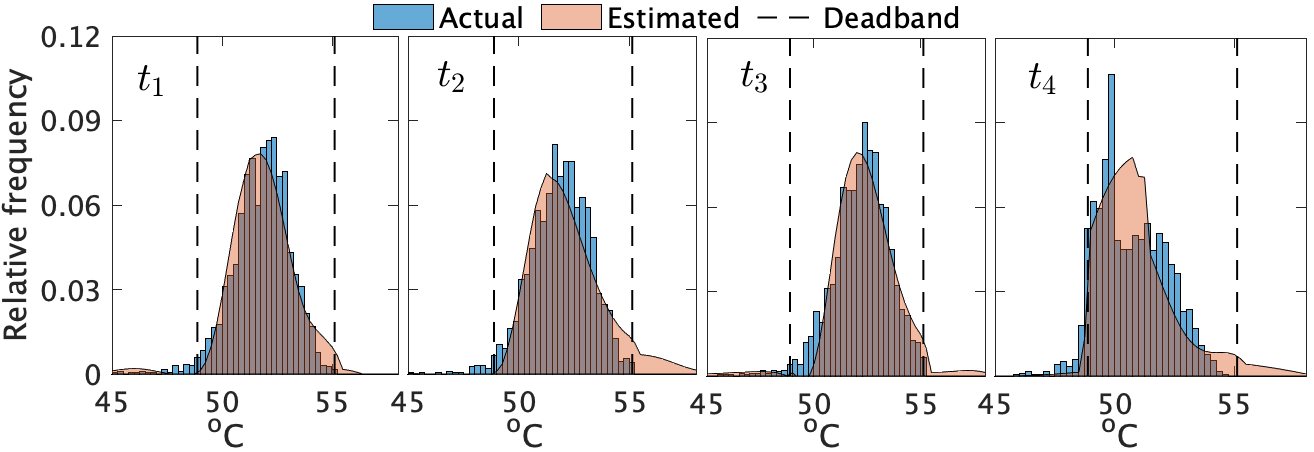}
    \caption{Normalized temperature distribution of $2,000$ EWHs at different times is shown in this plot. Estimated distribution from EKF in real-time (orange) matches closely with the actual distribution of EWHs (blue).}
    \label{fig:PVB_EKF_validation_Distribution}
\end{figure}

State estimation is illustrated on a fleet of $2,000$ homogeneous EWHs with parameters $\Theta_{\text{EWH}} = \{4.5,275,52, [48.9, 55.1]\}$ under PEM. The baseline power consumption is obtained from the optimization problem (\ref{eq:PEM_nonlin_dyn}) that results in $\beta_c^*[k] = 0.26$ and $P_{\text{nom}}^* = 2.35$ MW. Note that since EWHs cannot discharge power into the grid as batteries, then the control signal reduces to the scalar $\beta[k] = \beta_c[k]$. The maximum and minimum energy limits are $53.4^oC$ and $49.3^oC$ respectively. Figure~\ref{fig:PVB_EKF_validation} shows EWHs tracking a scaled and shifted AGC signal. The fleet's SoC is around nominal for about $2$ hours while tracking the AGC signal shifted to $P_{\text{nom}}^*$. EKF executed online accurately estimates the SoC in real-time from three measurements only; total power consumption, total number of request and opt-out rates. In the next hour, the reference is higher than $P_{\text{nom}}^*$ that causes EWHs to charge at an aggregate level. Finally, the fleet is discharging when the reference is below $P_{\text{nom}}^*$ for the last hour. This is also indicated by an increase in the number of requests and decrease in SoC. In addition to the SoC, EKF also predicts the temperature distribution of EWHs as shown in Fig.~\ref{fig:PVB_EKF_validation_Distribution}. Furthermore, this setup can also be used to tune the noise parameters of the EKF before real-world deployment. 

\section{Emulating DERs providing grid services} \label{sec:grid_service}
In this section, the real-time simulator is used to demonstrate that PEM resources can be used to manage tie-line mismatches in the grid under uncertain solar generation. Other grid services such as energy arbitrage, peak-load reduction etc. can also be tested but not shown in this paper due to space limitations. 

\subsection{Dispatching flexible resources in the grid}
\begin{figure}[ht]
    \centering
    \includegraphics[width=\columnwidth]{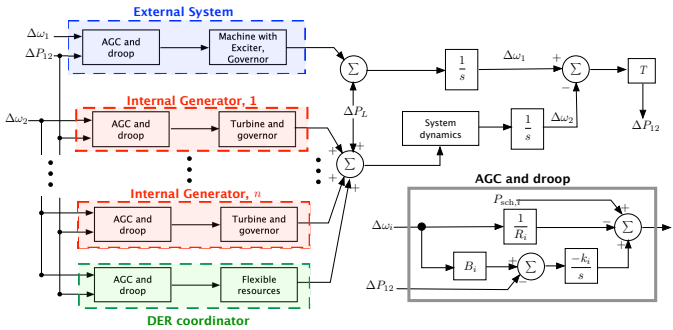}
    \caption{Control schematic for the VELCO System, including internal and external generation and DER fleet. This dynamic grid model is simulated with OPAL-RT's ePHASORSIM.}
    \label{fig:AGC_ControlDiagram1}
\end{figure}

Grid operators may pay high penalties for rescheduling generators or importing power through tie-lines to balance mismatches between supply and demand~\cite{Arnold2011PSCC}. These power mismatches can be balanced by controlling flexible resources. Primary frequency regulation (speed-droop) on each generator and flexible resources can stabilize the power system with a steady state frequency deviation from the desired system frequency depending on the droop characteristic and frequency sensitivity~\cite{Amini2016ISGT}. Furthermore, automatic generation control (AGC) is included to drive the steady state frequency deviation to zero. Figure~\ref{fig:AGC_ControlDiagram1} shows the generalized control diagram for the system considered in this paper, which is an adapted version of the tie line control from \cite{Kundur1994PowerSystemBook}. A linear combination of frequency errors and change in imported power through tie-lines from their scheduled contract basis is used as an error signal called the area control error (ACE). AGC acts as a secondary control using an integral controller that sends out control signals to generators and the DER coordinator to reduce ACE to zero in steady state. For the purposes of this work, only two areas are used for simplicity, while being effective enough to demonstrate the flow of power between different areas. The first area is referred to as the ``internal" area, representing the state of Vermont, whereas, the ``external" area is used to represent the import and export of power from and to the Vermont transmission system. Furthermore, AGC gains of the DER coordinator are designed so that the contribution of DERs in frequency regulation decreases with the increase in the fleet's SoC. Next, the details of the grid used in case studies is described.

\subsubsection{VELCO Test System}\label{Subsec:VELCOtestcase}
Vermont electric power company (VELCO) is Vermont's transmission system operator and provided the transmission grid data used to develop a realistic power system on which validation is performed. The ePHASORSIM model used in the simulator is then developed that consists of $161$ buses, $135$ branches, $89$ transformers and $22$ generation units that are lumped into two local generators labeled 1 and 2 respectively in Table~\ref{table:Grid_DER_param}. Furthermore, power is imported and exported through tie-line interconnects and power imports are modeled as a lumped constant-power generator called external generator. The ePHASORSIM model is initialized based on VELCO data and AC load flow with reactive power limits on generators and a single slack bus. The generators are modeled in ePHASORSIM using the $6$-th order synchronous machine model. The exciter is modeled using the IEEE type ACA4 with the time constant $0.01$ and the overall gain $200$. The droop parameters ($R$), bias factors ($B$) and the integral gain of the generators as well as the DER fleet have been tuned according to their respective capacities. These parameters are given in Table~\ref{table:Sys_param} and an illustrative example using this model is presented next.

\begin{table}[h]
\centering
\caption{Grid and DER parameters}
\label{table:Grid_DER_param}
\begin{tabular}{ l c l c}
\toprule
\multicolumn{4}{c}{Vermont Grid Parameters}\\ 
\midrule
No. of buses/branches & 161/223 & Total load & 609 MW \\
Total renewables & 305 MW & Cap. of external gen. & 240 MW \\
Cap. of local gen. 1 & 130 MW & Cap. of local gen. 2 & 35 MW \\
Cap. of bulk battery  & 45 MWh & Nominal DER load & 4.68 MW\\
\bottomrule
\end{tabular}
\end{table}
\begin{table}[h]
\centering
\caption{Droop and AGC parameters}
\label{table:Sys_param}
\begin{tabular}{l c c c c}
\toprule
Parameter & External & Local 1 & Local 2 & DER\\
\midrule
$R$ & 33 & 33 & 33 & 20\\
$B$ & 1 & 1 & 1 & 1\\
$k$ & 1 & 1 & 1 & 5.1\\
$P_{\text{gen}}^{\text{sch}}$ (MW) & 218 & 86.2 & 26.6 & 4.68\\
\bottomrule
\end{tabular}
\end{table}

\subsection{Illustrative example: accounting for uncertainty in solar}
\begin{figure}[ht]
    \centering
    \includegraphics[width=\columnwidth]{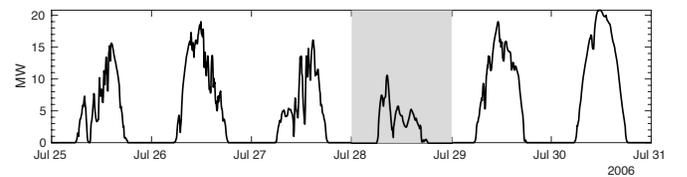}
    \caption{Variation in solar generation over several days \cite{NREL:SolarDataSet2006}. Case studies in this paper focuses on July $28$, highlighted in grey color.}
    \label{fig:SolarVariation}
\end{figure}

Consider the scenario in which Vermont has time-varying distributed solar generation and flexible load consisting of $4,000$ homogeneous EWHs. The base load of the fleet is $4.68$ MW according to the nominal response of PEM described earlier and the corresponding SoC limits are given in Table~\ref{table:Grid_DER_param}. Furthermore, Vermont has access to a bulk battery that can also be used. The objective of this bulk battery is to show that the DER coordinator can effectively co-optimize other resources along with flexible resources. Simulated solar power data for Vermont from \cite{NREL:SolarDataSet2006} is used as shown in Figure.~\ref{fig:SolarVariation} for several days in July, 2006. The results presented in this paper focuses on a single day, July 28. 

\begin{figure}[ht]
    \centering
    \includegraphics[width=\columnwidth]{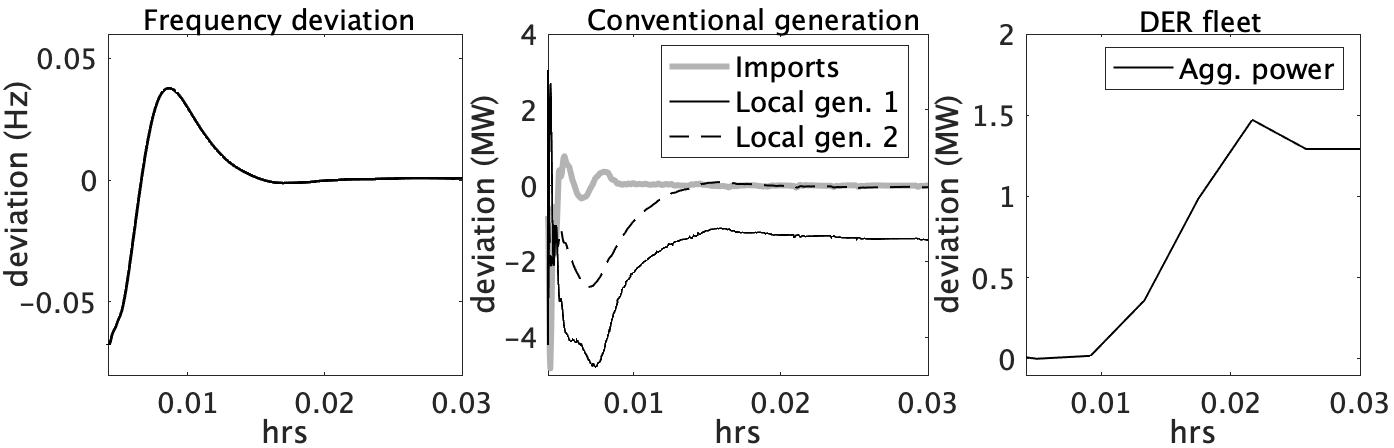}
    \vspace*{-2em}
    \caption{ The initial response of the system at the start of the simulation is shown in this figure. System frequency deviation returns to zero as the AGC dynamically adjusts local generators and DER fleet's power output to match increasing solar generation.}
    \label{fig:Results1}
\end{figure}

\begin{figure}[ht]
    \centering
        \includegraphics[width=\columnwidth]{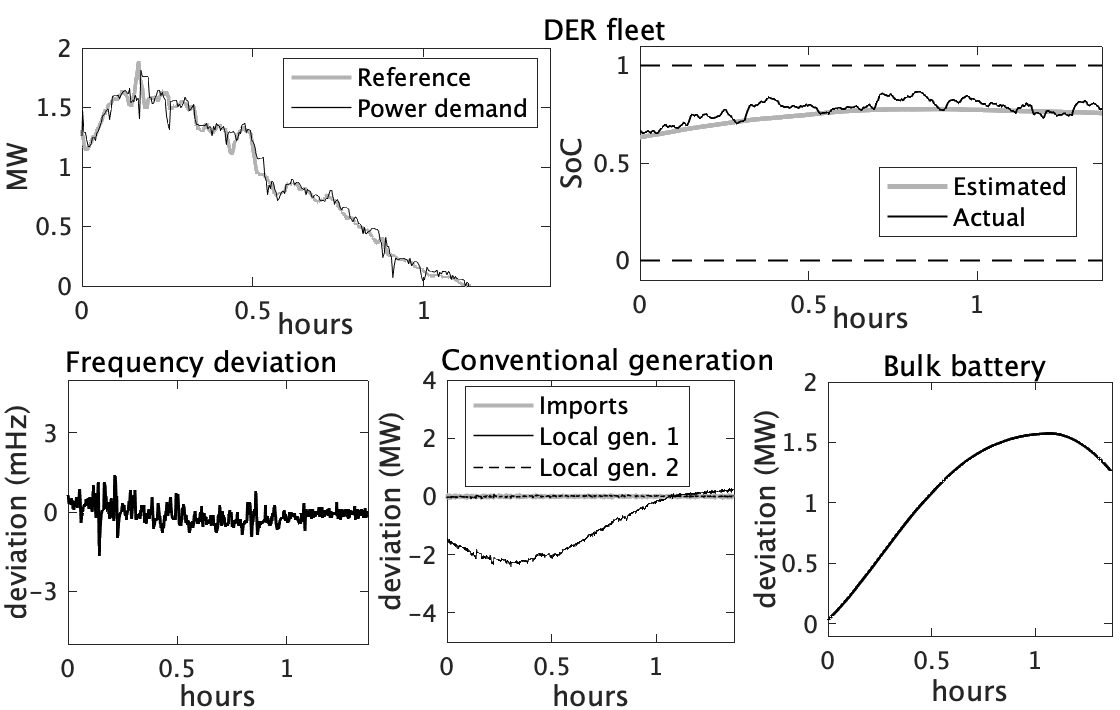}
        \vspace*{-2em}
    \caption{Top row shows the DER fleet tracking a reference signal generated by the AGC based on its SoC. The AGC is designed to account for the SoC of the fleet and as the SoC increases, the tracking power reference is lowered. Power reference and consumption is plotted in terms of deviation from the nominal consumption. Bottom row shows the frequency deviation of the system in mHz as well as the variation in power from conventional generation and total imported power. The imported power remains fixed at its scheduled value whereas the bulk battery and the less expensive local generation compensates for the excess solar generation.}
    \label{fig:Results2}
    \vspace*{-1.5em}
\end{figure}

The initial frequency deviation and the deviation in generator's power output as a result of increased solar generation is plotted in Fig.~\ref{fig:Results1} while the system performance over the next hour and a half is plotted in Fig.~\ref{fig:Results2}. AGC ensures that the power imported through the tie-line remains close to the scheduled value by dynamically modifying the setpoints of local generators, the bulk battery as well as the flexible EWHs under PEM coordinator as shown in Fig.~\ref{fig:Results2}. As a result, the EWH fleet starts to charge and its SoC increases. When the SoC gets closer to the upper limit, the AGC reduces the power setpoint of the PEM coordinator. It should also be noted that by using flexible resources, the more expensive imports remain fixed at their scheduled values whereas the less expensive EWHs, bulk battery and local generators compensate for the excess solar generation.

\section{Incorporating DER related practical limitations in the real-time simulator} \label{sec:delay}
This section shows that the real-time platform allows the coordinator to identify potential practical limitations in real-world deployment. Two cases are considered; ($i$) effect of measurement delays and ($ii$) delays introduced due to latency in the communication channel. The results presented here uses PEM for demonstration purposes, however, other schemes can also be implemented.

\subsection{Effect of Measurement delays}
Delays in the measurement of aggregate power $P_{\text{dem}}$ are studied first which can be delayed due to practical limitations. The coordinator then uses the most recent measurement of $P_{\text{dem}}$ to decide the number of requests to be accepted. It should be emphasized here that the coordinator is making these decisions in real-time and in a sequential manner. Recall that a DER once allowed to consume a packet will either charge or discharge for the entire duration of the packet length. Delays in the power measurement directly influences the performance of the DER coordinator since authorizing incorrect number of packets can not be cancelled under PEM. This is because the coordinator is unaware of the identity of the DER associated with each request that further ensures privacy of the participating DER. 

\begin{figure}[ht]
    \centering
    \includegraphics[width=\columnwidth]{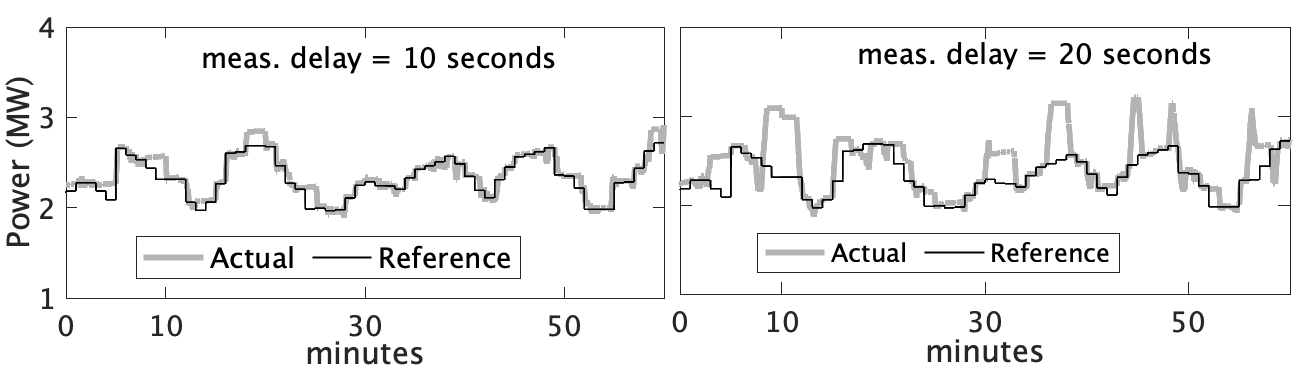}
    \vspace*{-2em}
    \caption{This plots shows the effect of measurement delays in $P_{\text{dem}}$ on tracking performance of DERs. The right plot is for the case when the average measurement delay is $20$ seconds resulting in RMS tracking error of $60$kW. Similarly, the left plot shows the case in which $P_{\text{dem}}$ measurements are delayed by $30$ seconds on average. RMS tracking error in this case is $160.6$kW.}
    \label{fig:MeasDelay}
\end{figure}

To study the effect of delays, it is assumed that the probability of a $P_{\text{dem}}$ measurement being delayed is $10\%$. Delay time itself is distributed according to a normal distribution with mean $\{20, 30, 60\}$ seconds and a standard deviation of $2$ seconds. Fig.~\ref{fig:MeasDelay} shows that a $20$ second average delay introduces an RMS tracking error of about $60$kW which translates to $2.5\%$ with respect to the baseline power consumption. The maximum RMS error of $15\%$ is observed in the extreme case of $60$ seconds. In reality, delays are usually less than $20$ seconds and Fig.~\ref{fig:MeasDelay} shows that PEM resources can deliver excellent tracking in real-time even in the presence of delays. However, in PEM the coordinator can also make use of the packet request mechanism to enhance the tracking performance as discussed next. 

\subsubsection{Constructing real-time power demand of DER fleet}
The inherent packet based mechanics in PEM allows the coordinator to approximate the aggregate power consumption of the fleet in real-time. Consider first the charging requests and recall that accepting a request corresponds to an increase in the aggregate power $P_{\text{dem}}$ by $P_{1}$, whereas, packet expiration corresponds to a decrease in power by $P_{1}$. Therefore, at any time instant, $P_{\text{dem}}$ of the fleet increases according to the total number of DERs authorized by the coordinator to charge and the total number of DERs opting out. Both these quantities are known to the coordinator as mentioned in section~\ref{sec:PEM_prelim}. In a homogeneous population of DERs, $P_{1}$ is the same for each packet. For heterogeneous DERs, an additional rated power field in the request can be added without extra communication overheard. In order to keep tracking of expiring packets, the coordinator associates each packet with a timer equal to the packet length $\delta_1$ as mentioned in section~\ref{subsec:PEM_implement}. As the timer expires, the coordinator models that the associated DER has transitioned to standby and adjusts $P_{\text{dem}}$ accordingly. This packet based timer mechanism allows the coordinator to construct $P_{\text{dem}}$ when measurements are delayed.

\begin{figure}[ht]
    \centering
    \includegraphics[width=\columnwidth]{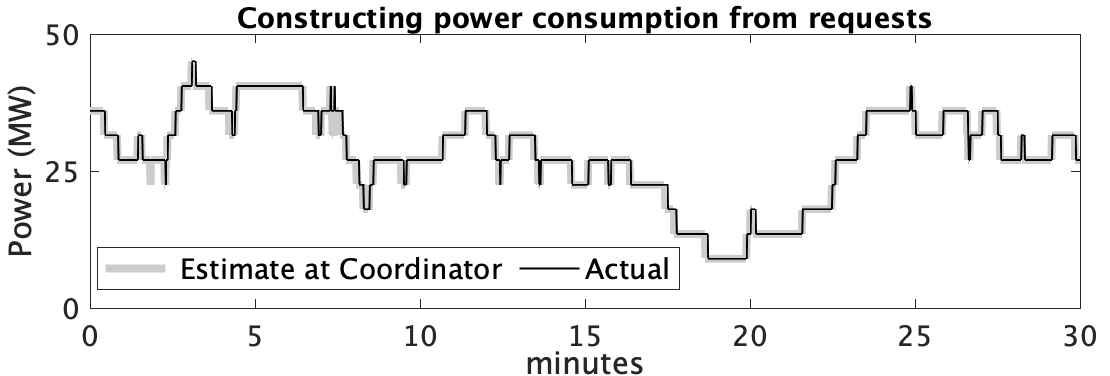}
    \vspace*{-2em}
    \caption{This figure shows that the DER coordinator can use the requests and a timer mechanism to construct power consumption estimate and is demonstrated here for $20$ EWHs.}
    \label{fig:ConstructPower}
\end{figure}

The plot in Fig.~\ref{fig:ConstructPower} shows that the estimated power matches that of the true aggregate power of the fleet ($P_{\text{dem}}$). Only $20$ EWHs are considered to demonstrate the accuracy of these measurements. These concepts are then easily extended to the case of discharging packets. DER coordinator operating under PEM is therefore equipped with a built-in feedback mechanism in terms of requests that allows tight tracking performance and accurate SoC estimation observed in the earlier section.

\vspace*{-2em}
\subsection{Effect of input delays}
\begin{figure}[ht]
    \centering
    \includegraphics[width=\columnwidth]{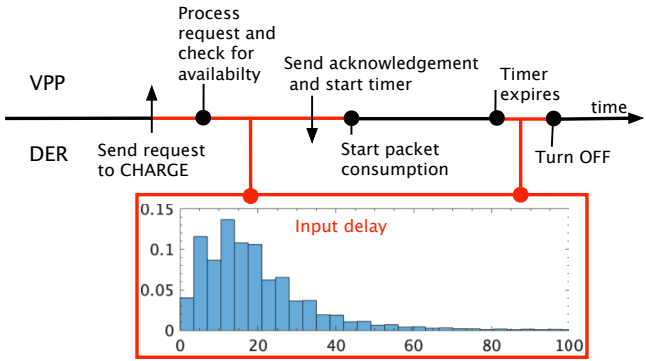}
    \caption{(Top) Sequence of events that occur between the aggregator and the DERs in PEM's request mechanism, (bottom) measured input delay between request acceptance and packet consumption in PEM.}
    \label{fig:InputDelayMeasurements}
\end{figure}

Input delays are incurred due to the time between the DER coordinator sending a charge/discharge authorization and the corresponding DER switching its state. To understand the source of such delays, consider the sequence of events in PEM's request mechanism, conceptually represented in Fig.~\ref{fig:InputDelayMeasurements}. The DER sends a request to consume a packet to the coordinator which then responds with an acknowledgement and initiates a timer. Whereas, the DER waiting for the response, starts to consume the packet only after the acknowledgement has arrived. Furthermore, when the timer at the coordinator expires, it assumes that $P_{\text{dem}}$ should be reduced. However, it is possible that the DER's own dynamics might extend this packet consumption time and lead to an additional delay. For the developed cyber-enabled real-time simulator, the combination of both these delays is distributed as shown in Fig.~\ref{fig:InputDelayMeasurements} with a mean of $2$ms for $2,000$ EWHs. RMS error between the actual power consumption and the estimated power consumption is $13.06$kW. In real-world deployment of \cite{Adil2019ISGT}, these delays are observed to be $8$ms on average. To study the effect of increased measurement delays on the DER coordinator's estimates of power consumption, $8$ms input delays are artificially injected. This results in RMS error of $35.4$kW which is only about $1.7\%$ of the nominal power consumption showing good tracking performance.
\vspace*{-1.5em}
\section{Conclusion} \label{sec:concl}
This manuscript presented a real-time grid-and-DER co-simulation platform that enables and characterizes performance evaluation of large-scale DER coordination schemes in a realistic setting. Importantly, to support validation efforts and lower barriers to real-world DER program deployments, this platform captures salient cyber and physical limitations, including communication systems, device behaviors, and grid challenges. Moreover, the co-simulator is agnostic to the type of coordination scheme and allows real-time implementation of model-based state-estimation and control algorithms. Illustrative results from the co-simulation platform are presented for the DER demand dispatch scheme called PEM. PEM has a unique request-response mechanism that coordinates DERs via bi-directional DER-to-coordinator communications, which makes it well suited for exploring the capability of the real-time grid-and-DER co-simulation platform.

Future work seeks to implement and compare different centralized and distributed coordination schemes in terms of performance and QoS guarantees. Furthermore, since different types of DERs operating under the same coordinator are suitable for several applications such as peak-load reduction, frequency regulation etc., therefore, we are interested in application specific grouping and validation of diverse DER fleets.
\vspace*{-1em}
\section*{Acknowledgment}
The authors would like to thank Andrew Klem, Prof. Jeff Frolik from University of Vermont and Prof. Sumit Paudyal from Florida International University for valuable discussions on PEM based real-time grid simulations and experiments. Furthermore, we are grateful to Ms. Bernadette Fernandes and Mr. Chris Root of VELCO for insightful discussions and providing transmission and bus data of the Vermont system to create a realistic case study.


\begin{thebibliography}{10}

\bibitem{AminiPSCC2018}
M.~{Amini} and M.~{Almassalkhi}, ``Trading off robustness and performance in
  receding horizon control with uncertain energy resources,'' in {\em 2018
  Power Systems Computation Conference (PSCC)}, pp.~1--7, June 2018.

\bibitem{schweppe1980homeostatic}
F.~C. Schweppe, R.~D. Tabors, J.~L. Kirtley, H.~R. Outhred, F.~H. Pickel, and
  A.~J. Cox, ``Homeostatic utility control,'' \emph{IEEE Transactions on Power Apparatus and Systems}, no.~3, pp. 1151--1163, 1980.

\bibitem{Morgan:1979}
M.~Morgan and S.~Talukdar, ``Electric power load management: Some technical,
  economic, regulatory and social issues,'' {\em Proceedings of the IEEE},
  vol.~67, no.~2, pp.~241 -- 312, 1979.

\bibitem{malhame1990AAP}
R.~Malham\'e, ``A jump-driven markovian electric load model,'' \emph{Advances
  in Applied Probability}, vol.~22, no.~3, p. 564--586, 1990.

\bibitem{Callaway2009ECM}
D.~S. Callaway, ``Tapping the energy storage potential in electric loads to
  deliver load following and regulation, with application to wind energy,''
  {\em Energy Conversion and Management}, vol.~50, no.~5, pp.~1389 -- 1400,
  2009.

\bibitem{Meyn2015TAC}
S.~P. Meyn, P.~Barooah, A.~Busic, Y.~Chen, and J.~Ehren, ``{Ancillary Service to the Grid Using Intelligent Deferrable Loads},'' \emph{IEEE Transactions on Automatic Control}, vol.~60, no.~11, pp. 2847--2862, 2015.

\bibitem{Mathieu:2013tt}
J.~L. Mathieu, M.~Kamgarpour, J.~Lygeros, and D.~S. Callaway, ``{Energy
  arbitrage with thermostatically controlled loads},'' in {\em European
  Conference on Circuit Theory and Design}, pp.~2519--2526, July 2013.

\bibitem{Luminita2014IFAC}
L.~C. Totu and R.~Wisniewski, ``Demand response of thermostatic loads by
  optimized switching-fraction broadcast,'' {\em IFAC Proceedings Volumes},
  vol.~47, no.~3, pp.~9956 -- 9961, 2014.
\newblock 19th IFAC World Congress.

\bibitem{almassalkhi2018BookChapter}
M.~Almassalkhi, L.~A. Duffaut~Espinosa, P.~D.~H~Hines, J.~Frolik, S.~Paudyal, and
  M.~Amini, \emph{Asynchronous coordination of distributed energy resources with Packetized energy management}.\hskip 1em plus 0.5em minus 0.4em\relax
  New York, NY: Springer New York, 2018, pp. 333--361.

\bibitem{DuffautEspinosa:2018PSCC}
L.~A. {Duffaut Espinosa}, M.~Almassalkhi, P.~Hines, and J.~Frolik, ``System properties of packetized energy management for aggregated diverse resources,'' \emph{Power Systems Computation Conference}, June 2018.

\bibitem{Duffaut-et-al_2019a}
L.~A. Duffaut~Espinosa and M.~Almassalkhi, ``A packetized energy management
  macromodel with quality of service guarantees for demand-side resources.''
  \emph{IEEE Transactions on Power Systems}, vol.~35, no.~5, pp. 3660--3670, 2020.

\bibitem{Duffaut-et-al_2019b}
L.~A. {Duffaut Espinosa}, A.~Khurram, and M.~Almassalkhi, ``Reference-tracking
  control policies for packetized coordination of heterogeneous DER
  populations.,'' {\em IEEE Transactions on Control Systems Technology}, pp.~1--17, 2020.

\bibitem{Serpanos2018Computer}
D.~{Serpanos}, ``The cyber-physical systems revolution,'' {\em Computer},
  vol.~51, pp.~70--73, March 2018.

\bibitem{Xin2015TSG}
S.~{Xin}, Q.~{Guo}, H.~{Sun}, B.~{Zhang}, J.~{Wang}, and C.~{Chen},
  ``Cyber-physical modeling and cyber-contingency assessment of hierarchical
  control systems,'' {\em IEEE Transactions on Smart Grid}, vol.~6,
  pp.~2375--2385, 2015.

\bibitem{Ayar2017IET}
M.~{Ayar}, R.~D. {Trevizan}, S.~{Obuz}, A.~S. {Bretas}, H.~A. {Latchman}, and
  N.~G. {Bretas}, ``Cyber-physical robust control framework for enhancing
  transient stability of smart grids,'' {\em IET Cyber-Physical Systems: Theory
  Applications}, vol.~2, no.~4, pp.~198--206, 2017.

\bibitem{Korukonda2018IET}
M.~P. {Korukonda}, S.~R. {Mishra}, K.~{Rajawat}, and L.~{Behera}, ``Hybrid
  adaptive framework for coordinated control of distributed generators in
  cyber-physical energy systems,'' {\em IET Cyber-Physical Systems: Theory
  Applications}, vol.~3, no.~1, pp.~54--62, 2018.

\bibitem{NREL_CPS:EUROCON2019}
A.~{Pratt}, M.~{Baggu}, F.~{Ding}, S.~{Veda}, I.~{Mendoza}, and E.~{Lightner},
  ``A test bed to evaluate advanced distribution management systems for modern
  power systems,'' in {\em IEEE EUROCON 2019 -18th International Conference on
  Smart Technologies}, pp.~1--6, 2019.

\bibitem{Hahn:TSG2013}
A.~{Hahn}, A.~{Ashok}, S.~{Sridhar}, and M.~{Govindarasu}, ``Cyber-physical
  security testbeds: Architecture, application, and evaluation for smart
  grid,'' {\em IEEE Transactions on Smart Grid}, vol.~4, no.~2, pp.~847--855,
  2013.

\bibitem{UIUC_CPS:Informs2009}
D.~Nicol, C.~Davis, and T.~Overbye, ``A virtual power system testbed for
  cyber-security decision support,'' in {\em Proceedings of the 2009 INFORMS
  simulation society workshop on simulation: At the interface of modeling and
  anaylsis}, pp.~62--66, 2009.

\bibitem{Ledva:2017ws}
G.~S. Ledva, E.~Vrettos, S.~Mastellone, G.~Andersson and J.~Mathieu, ``{Managing communication delays and model error in demand response for frequency regulation},'' in {\em IEEE Transactions on Power Systems}, vol.~33, no.~2, p.~1299--1308, 2018.

\bibitem{Amini2019PESGM}
M.~Amini, A.~Khurram, A.~Klem, M.~Almassalkhi, and P.~D. Hines, ``A
  model-predictive control method for coordinating virtual power plants and
  packetized resources, with hardware-in-the-loop validation,'' in {\em
  IEEE Power Energy Society General Meeting (PESGM)}, pp.~1--5, 2019.

\bibitem{Adil2019ISGT}
K.~Desrochers, A.~Khurram, M.~Amini, A.~Giroux, F.~Wallace, J.~Slinkman,
  V.~Hines, M.~Almassalkhi, and P.~Hines, ``{Real-world, full-scale validation of power balancing services from Packetized virtual batteries},'' in
  \emph{Conference on Innovative Smart Grid Technologies {N}orth America}, pp.~1--5, 2019.

\bibitem{Arnold2011PSCC}
M.~Arnold and G.~Andersson, ``Model predictive control of energy storage
  including uncertain forecasts,'' in {\em Power Systems Computation Conference}, 2011.

\bibitem{Amini2016ISGT}
M.~{Amini} and M.~{Almassalkhi}, ``Investigating delays in frequency-dependent
  load control,'' in {\em IEEE Innovative Smart Grid Technologies - Asia
  (ISGT-Asia)}, pp.~448--453, Nov. 2016.

\bibitem{Kundur1994PowerSystemBook}
P.~Kundur, N.~J. Balu, and M.~G. Lauby, {\em Power system stability and
  control}, vol.~7.
\newblock McGraw-hill New York, 1994.

\bibitem{NREL:SolarDataSet2006}
{National Renewable Energy Laboratory} (NREL), ``Solar power data for
  integration studies,'' 2006. [Online]. Available: \url{https://www.nrel.gov/grid/solar-power-data.html}.

\end{thebibliography}
\end{document}